\documentclass[journal]{./IEEEtran}
\usepackage{ifpdf}
\usepackage[latin1]{inputenc}
\usepackage{graphicx}
\usepackage{epsfig}
\usepackage{amssymb,amsmath,array}
\usepackage{epstopdf}

\ifCLASSINFOpdf
\else
\fi
\begin{document}
%
\title{DEPFET active pixel detectors for a future linear $e^+e^-$ collider}
%
%
%
\author{{ \bf The DEPFET collaboration} \\
  (www.depfet.org) \\ 
O. Alonso, R. Casanova, A. Dieguez \\ 
 {\em Universitat de Barcelona, Spain} \\
J. Dingfelder, T. Hemperek, T. Kishishita, T. Kleinohl,
M. Koch, H. Kr\"uger, M. Lemarenko, F. L\"utticke, \mbox{C. Marinas}, M. Schnell, N. Wermes \\
{\em Bonn University, Germany} \\
 A. Campbell, T. Ferber, C. Kleinwort, C. Niebuhr, Y. Soloviev, 
M. Steder, R. Volkenborn, \mbox{S. Yaschenko} \\
{\em Deutsches Elektronen-Synchrotron, Hamburg, Germany}
\\
P. Fischer, C. Kreidl, I. Peric, J. Knopf, M. Ritzert \\   
{\em Heidelberg University, Germany} \\
E. Curras, A. Lopez-Virto, D. Moya, I. Vila \\
{\em Instituto de F\'isica de Cantabria (CSIC-UC), Santander, Spain} \\ 
M. Boronat, D. Esperante, J. Fuster, I. Garcia Garcia, C. Lacasta, A. Oyanguren, P. Ruiz, G. Timon, M. Vos$^{*}$ \\
 {\em Instituto de F\'isica Corpuscular (UVEG/CSIC), Valencia, Spain}  \\
T. Gessler, W. K\"uhn, S. Lange, D. M\"unchow, B. Spruck \\
 {\em Giessen University, Germany} \\
 A. Frey, C. Geisler, B. Schwenker, F. Wilk  \\ 
 {\em G\"ottingen University, Germany}  \\
T. Barvich, M. Heck, S. Heindl, O. Lutz, Th. M\"uller, C. Pulvermacher, H.J. Simonis, T. Weiler  \\ 
 {\em KIT Karlsruhe, Germany}\\
T. Krausser, O. Lipsky, S. Rummel, J. Schieck, T. Schl\"uter \\
{\em Ludwig-Maximilians-University, Munich, Germany} \\
K. Ackermann, L. Andricek, V. Chekelian, V. Chobanova, J. Dalseno, C. Kiesling, C. Koffmane, L. Li Gioi, \mbox{A. Moll}, H. G. Moser, F. M\"uller, E. Nedelkovska, J. Ninkovic, S. Petrovics, K. Prothmann, R. Richter, A. Ritter, \mbox{M. Ritter}, F. Simon, P. Vanhoefer, A. Wassatsch  \\ 
 {\em MPI Munich, Germany}  \\
Z. Dolezal, Z. Drasal, P. Kodys, P. Kvasnicka, J. Scheirich  \\ 
{\em Charles University, Prague, Czech Republic} \\ 
\thanks{$^{*}$ Corresponding author. E-mail: marcel.vos@ific.uv.es}
}
%
%
\markboth{Transactions on Nuclear Science,~Vol.~6, No.~1, September~2010}%
{Shell \MakeLowercase{\textit{et al.}}: DEPFET active pixel detectors for a future linear $e+e-$ collider}
%
\maketitle

\begin{abstract}
The DEPFET collaboration develops highly granular, ultra-transparent active 
pixel detectors for high-performance vertex reconstruction at future
collider experiments. The 
characterization of detector prototypes has proven that the key principle, 
the integration of a first amplification stage in a detector-grade sensor 
material, can provide a comfortable signal to noise ratio of over 40
for a sensor thickness of 50-75 $\mathrm{\mathbf{ \mu m}}$. 
ASICs have been designed and produced to operate a DEPFET pixel 
detector with the required read-out speed. 
A complete detector concept is being developed, including solutions for 
mechanical support, cooling and services. In this paper the status of DEPFET 
R \& D project is reviewed in the light of the requirements of the vertex 
detector at a future linear $\mathbf{e^+ e^-}$ collider.
\end{abstract}


%
\IEEEpeerreviewmaketitle

\section{Introduction}

Experiments at a future linear $e^+ e^-$ collider~\cite{ilcrdr,cliccdr} (LC) 
requires extremely precise reconstruction of the reaction products 
to perform precision physics programs to study the electroweak symmetry 
breaking mechanism and physics beyond the Standard Model. 
Key figures of merit for the detector performance, such as the jet energy
resolution, momentum resolution for charged tracks and the vertexing
capabilities of the experiment, must be improved significantly with respect
to the state-of-the-art detectors realized in the LHC experiments.
A worldwide detector \mbox{R \& D} effort is ongoing to 
fully satisfy the challenging requirements.

Finely segmented solid state detectors are crucial for the reconstruction 
of the trajectory of charged particles in modern collider experiments. 
In active pixel detectors the small signal generated when charged particles
traverse a thin layer of Silicon is amplified in the sensor itself.
Over the last decade an international collaboration has developed the DEPFET 
(Depleted Field Effect Transistor~\cite{lutzkemmer}) concept, where a FET
is integrated in the active sensor to amplify the signal.
These DEPFET structures present interesting 
possibilities for a number of applications. The excellent signal to noise
(S/N) ratio that can be achieved has led to applications such as
space-based X-ray Astronomy missions and X-ray detection in the European
XFEL~\cite{Lutz:2007zza}. It also allows for a strong reduction 
of the material budget of position-sensitive devices for charged
particle detection at collider experiments. The proposal of a DEPFET vertex
detector for a future linear $e^+e^-$ 
collider~\cite{Richter:2003dn,Fischer:2002si} dates back to 2002. 
The DEPFET collaboration has since shown that finely segmented devices with 
large in-pixel gain can indeed be constructed and operated. Read-out
and control ASICs have been designed and produced, and a novel ladder 
design with excellent thermo-mechanical properties has been developed.
A fully engineered vertex detector design, including all supports
and services, is being developed for 
the Belle~II experiment~\cite{b2tdr}.

Recent descriptions of the DEPFET active pixel detector project are found 
in References~\cite{Vos:2009zz} and~\cite{Moser:2007zz}. 
The Belle~II technical
design report~\cite{b2tdr} provides a very complete description 
of the DEPFET-based vertex detector of that experiment 
(see also~\cite{Simon:2010mg, dolezal}). However, the most recent document
reviewing DEPFET progress towards a vertex detector for a linear
collider~\cite{VXDreview} was prepared for the ILC \mbox{R \& D} panel in 2007. 
The present paper aims to fill this gap, reviewing the major milestones of the
DEPFET project since then in the light of a linear $e^+ e^-$ collider at 
the energy frontier.

The requirements on an LC vertex detector are briefly summarized in 
Section~\ref{sec:lcrequirements}. The DEPFET concept is  
reviewed in Section~\ref{sec:concept} and a ladder design for an LC
vertex detector based on DEPFETs is schematically presented in 
Section~\ref{sec:thinning}. The following Sections highlight 
recent progress in several key areas, such as: sensor production 
(Section~\ref{sec:sensors}), ASIC development and 
performance (Section~\ref{sec:asics}), and power consumption 
and cooling (Section~\ref{sec:cooling}). In Section~\ref{sec:tb}
results from the characterization of prototypes are presented.
The results are summarized in Section~\ref{sec:conclusions}.

\section{Requirements on an LC vertex detector}
\label{sec:lcrequirements}

Detector concept studies for future linear colliders have established a number
of challenging performance goals based on the analysis of benchmark 
channels and an evaluation of the LC 
environment~\cite{cliccdr,ildloi,sidloi,rdrdetector,teslatdr}. 

The first layer of the LC vertex detector must cope with large irreducible 
backgrounds due to incoherent $e^+ e^-$ pair production. Long bunch trains are 
envisaged in the ILC Technical Design Report, with 1312 bunches separated 
by 554 ns~\footnote{These numbers have changed with respect to 
the LOI, that contemplated 2820 bunches with 337 ns spacing. However,
these changes are accompanied by a factor two increase in bunch current,
such that the background hit density per unit area and unit time 
remains virtually unaltered.}. A background hit density of up to
10 hits$/\mathrm{cm}^2/\mu\mathrm{s}$ in the innermost layers of the ILC
vertex detector at 15 mm from the interaction point require read out times 
of 50-100 $\mathrm{\mu s}$ and a highly granular detector~\cite{ildloi,sidloi}. 

The background hit rates scale approximately with the instantaneous luminosity, 
which is in turn expected to increase proportionally
with the center-of-mass energy of the machine. The read-out speed requirements
are therefore considerably more relaxed in the early low-energy stage 
(at 250 $-$ 350 GeV) than in the nominal 500 GeV stage and in the envisaged 
1 TeV upgrade.
CLIC envisages shorter trains of order 300 bunches with a 0.5 ns spacing.
Time stamping of single bunches in this environment requires devices 
with a read-out speed beyond the current state of the art, a challenge
that is not addressed in this Transaction.

The aim for the vertexing
capabilities of the LC detectors is often summarized with the following
requirement on the impact parameter resolution\footnote{The 
value of the material term is increased from 10 $\mathrm{\mu m}$ 
to 15 $\mathrm{\mu m}$ in the CLIC requirement as a consequence of the 
increased inner radius.} :
\begin{equation} 
  \Delta d_0 = 5 \/ [\mu \mathrm{m}] \oplus \frac{10 \/ [\mu \mathrm{m}] }{p \/ [GeV] \sin^{3/2}{\theta}}  
\label{eq:impactparameter}
\end{equation}
This goal respresent a considerable improvement over vertex detectors built at 
collider experiments to date; the constant term is better by a factor 2-4 than 
what was achieved at previous $e^+ e^-$ colliders and at the LHC. Achieving
the requirement for the second (material) term is even more challenging; 
it has to decrease by a factor 6-10 with respect to most 
previous experiments\footnote{The best material term so far was achieved 
by the SLD vertex detector with
 $\Delta d_0 = 9 [\mu \mathrm{m}] \oplus \frac{33 [\mu \mathrm{m}]}{p [\mathrm{GeV}] \sin^{3/2}{\theta}}$. 
This CCD based detector had 0.36\% of a radiation length per layer.}. 
Indeed, the requirement in Formula~\ref{eq:impactparameter} 
(together with the assumed inner radius of 15 mm) implies that the vertex 
detector must be built with a strict material budget of order 0.1\% of a 
radiation length per layer, roughly equivalent to 100 $\mathrm{\mu m}$ of 
Silicon.

To comply with the strict material budget, the detector concepts aim to keep
the (average) power consumption below 10 W for the entire vertex detector. 
With such a low power density (approximately 100 $\mathrm{mW/cm}^2$) no 
active cooling circuits are required. The key to achieve this goal is the
bunch structure of the LC machines. The 0.7 ms long bunch trains at the ILC are 
separated by intervals of 200 ms. In a pulsed power scheme, where the
detector power supply follows the machine duty cycle as closely as possible, 
a reduction of the average power consumption by a factor of (at maximum) 
1/275 is possible. The CLIC bunch trains, with a length of just 156 ns, are 
separated by 20 ms.

The detectors must be operated in a strong magnetic field, ranging from 
3.5 to 5 Tesla in the different detector concepts.
The radiation levels in the innermost part of the tracker volume of an LC 
detector are relatively low compared to the LHC. 
The non-ionizing dose is of the order of
$\mathrm{10^{10} - 10^{11} n_{eq}/cm^2/yr}$, while the ionizing radiation 
amounts to less than $1$~kGy per year~\cite{ildloi,cliccdr}.  

\section{The Depleted Field Effect Transistor}
\label{sec:concept}

A brief introduction of the Depleted Field Effect 
Transistor is presented in the following. 
For a complete discussion the reader is referred to 
more detailed descriptions elsewhere~\cite{b2tdr,VXDreview}.

\begin{figure}
\centering
\includegraphics[width=\linewidth]{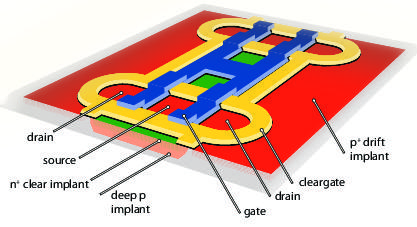}
\caption{Layout of the depleted Field Effect Transistor (DEPFET). 
The components of the FET - source, gate and drain - are indicated 
in the picture, as well as the clear contact and clear gate structure (the ring
structure around the FET). The drift structure depicted 
here is only present in the large pixels for Belle~II.}
\label{fig:depfet_structure}
\end{figure}

A schematic view of the DEPFET concept is shown in 
Figure~\ref{fig:depfet_structure}. 
A Field Effect Transistor (FET) is embedded in detector-grade Silicon.
An electric field is set up in the sensor to deplete it of free charge carriers
and to ensure fast signal collection with limited diffusion. 
The signal is collected on the {\em internal gate}, an $n^+$ implant 
immediately below the FET, where it modulates the drain current. 
After read-out the collected signal is removed from the internal gate 
by applying a voltage pulse to the clear contact.

With this structure a first amplification of the signal is achieved, 
that allows to reduce the active detector thickness to several tens of microns 
while maintaining a comfortable S/N ratio for minimum 
ionizing particles (MIPs). The gain $g_q = d I_{drain} / d q $ 
of this first stage (expressed in units of 
current per electron) is one of the crucial parameters of the DEPFET.

\section{DEPFET active pixel detectors}
\label{sec:thinning}
\begin{figure}[h!]
\centering
\includegraphics[width=\linewidth]{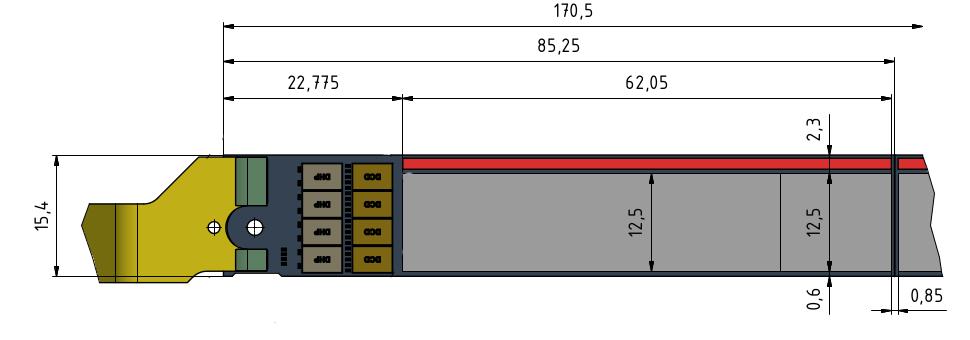}
\caption{Schematic ladder layout for the innermost layer of an LC vertex 
detector. The end-of-ladder area that houses the read-out ASICs is visible
on the leftmost side of the figure. The active area is indicated in 
light grey. The balcony that houses the control electronics is seen on
top of the picture, running along the full length of the sensor. Dimensions
are given in millimeters.}
\label{fig:schematicladder}
\end{figure}

A finely segmented active pixel sensor is achieved by introducing a 
matrix of very small DEPFET structures on the surface of the sensor. 
A schematic of a DEPFET ladder for the innermost layer of an LC vertex
detector is depicted in Figure~\ref{fig:schematicladder}.
The matrix of pixels is read out in a rolling shutter architecture. Control 
ASICs located on a narrow {\em balcony} that stretches along the length of the 
ladder address subsequent {\em rows} of pixels. Each {\em column} composed 
of approximately 1000 DEPFET pixels is read out by a single channel of 
a read-out chip located at the end of the sensor. 
As the shutter rolls over the matrix, each 
row of pixels\footnote{In practice, two or four rows are read out in
parallel to reduce the time required to read out a complete frame.} 
is switched on in turn. The frame time needed to read out the 
complete matrix is then given by the depth of the column multiplied by the time 
required to read out a single pixel. The auxiliary ASICs required to operate 
and read out the DEPFET sensors are discussed in more detail in 
Section~\ref{sec:asics}

To comply with the very tight material budget the sensor must be thinned 
and the material in support and services must be reduced to the bare minimum. 
The all-Silicon concept aims to achieve both by integrating active material, 
support structures and the high-density interconnect in a self-supporting 
ladder~\cite{VTXdepfetmech2,VTXdepfetmech}.

\begin{figure}[h!]
\centering
\includegraphics[width=\linewidth]{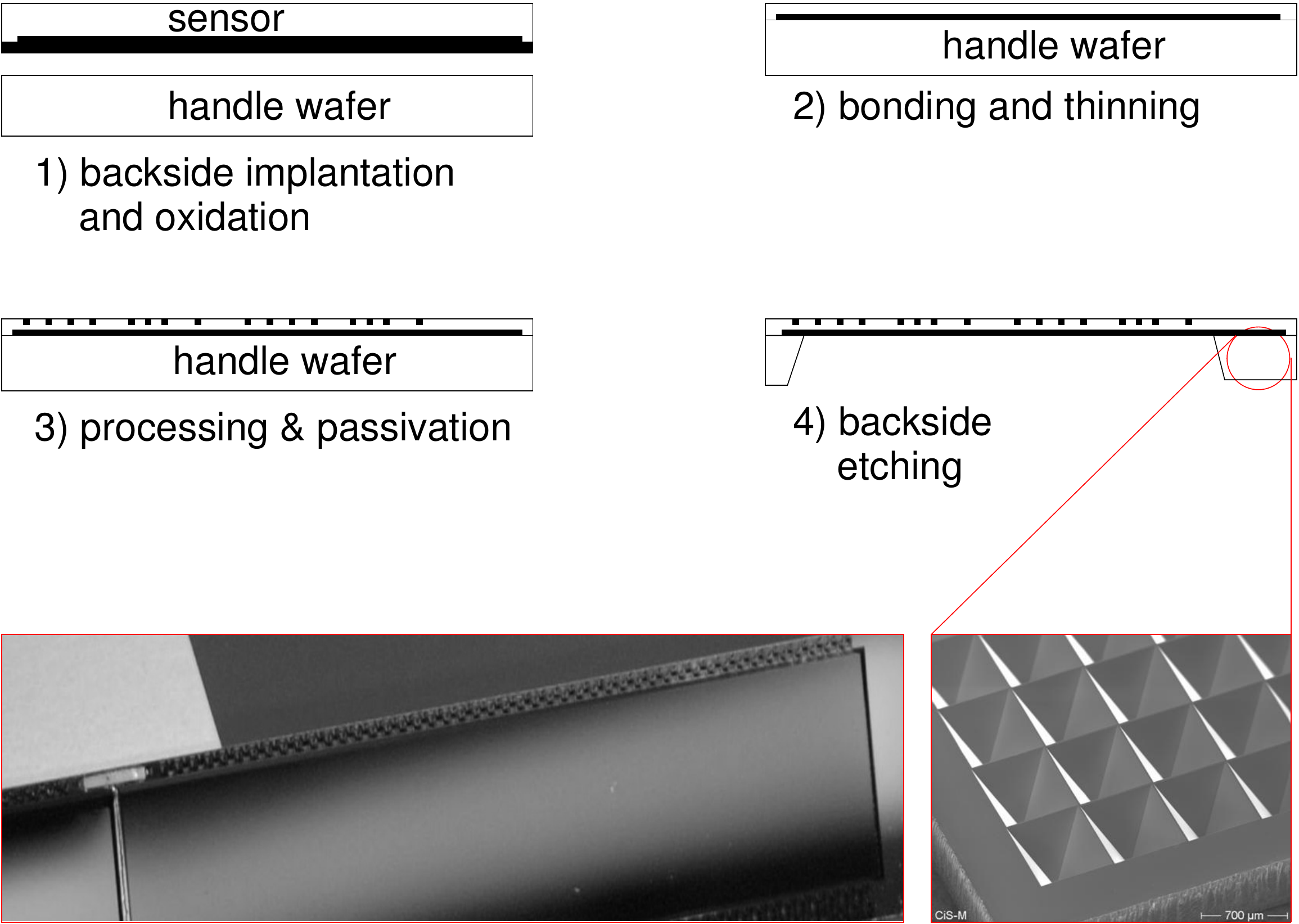}
\caption{Illustration of the most imporant steps in the creation of a thin all-Silicon ladder; (1) backside implants and oxidation of the sensor wafer; (2) bonding of the sensor and handle wafer and thinning of the former; (3) front-side
processing and passivation; (4) photolitographic thinning of the handle wafer, leaving support structures around the edges. A photograph of the ladder is also shown, as well as a scanning electron microscope image of a detail of the support structure around the edge. }
\label{fig:thinning}
\end{figure}

The most important process steps are illustrated in 
Figure~\ref{fig:thinning}. In the first step the backside implants are 
placed and the sensor wafer is oxidized. In the second step the sensor
wafer is bonded to a thick handle wafer; the sensor is 
ground to a thickness of 50 $\mathrm{\mu m}$ (75 $\mathrm{\mu m}$ for Belle~II).
In the third step the DEPFET processing is performed on the front side. 
Through a final photolithographic step (deep anisotropic etching) windows 
are opened in the second (support) wafer below the sensitive part of the 
sensor. The thicker Silicon around the edges of the sensor forms a support 
frame. A photo of the resulting ladder is shown in the same Figure. 
The material in the edges is further reduced in the same lithography step 
by introducing grooves. With the etching technique used complex structures
can be produced. A good example is shown in Figure~\ref{fig:thinning}, that
presents a scanning electron microscope image of an edge produced with this
technique. The all-Silicon ladders have excellent mechanical properties.
The all-Silicon mechanical concept is fully self-supporting and 
requires no external support structure over the length of the ladder. 
The use of a single material furthermore reduces 
the mechanical stress due to mismatching of thermal coefficients.


\begin{figure}[h!]
\centering
\includegraphics[width=0.45\linewidth]{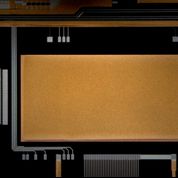}
\includegraphics[width=0.45\linewidth]{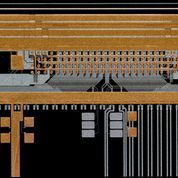}
\caption{Photographs of part of the high-density interconnections in Copper and Aluminium layers on the DEPFET ladder.}
\label{fig:emcm}
\end{figure}
Auxiliary detector components such as the control and read-out ASICs are 
integrated onto the sensor using bump-bonding. 
Power and signal lines are routed on an additional metal layer on the sensor, 
thus obviating the need for a separate high-density interconnect cable.
A close-up image of part of the high-density interconnections on a Belle~II
ladder is shown in Figure~\ref{fig:emcm}.

\begin{figure}[h!]
\centering
\includegraphics[width=\linewidth]{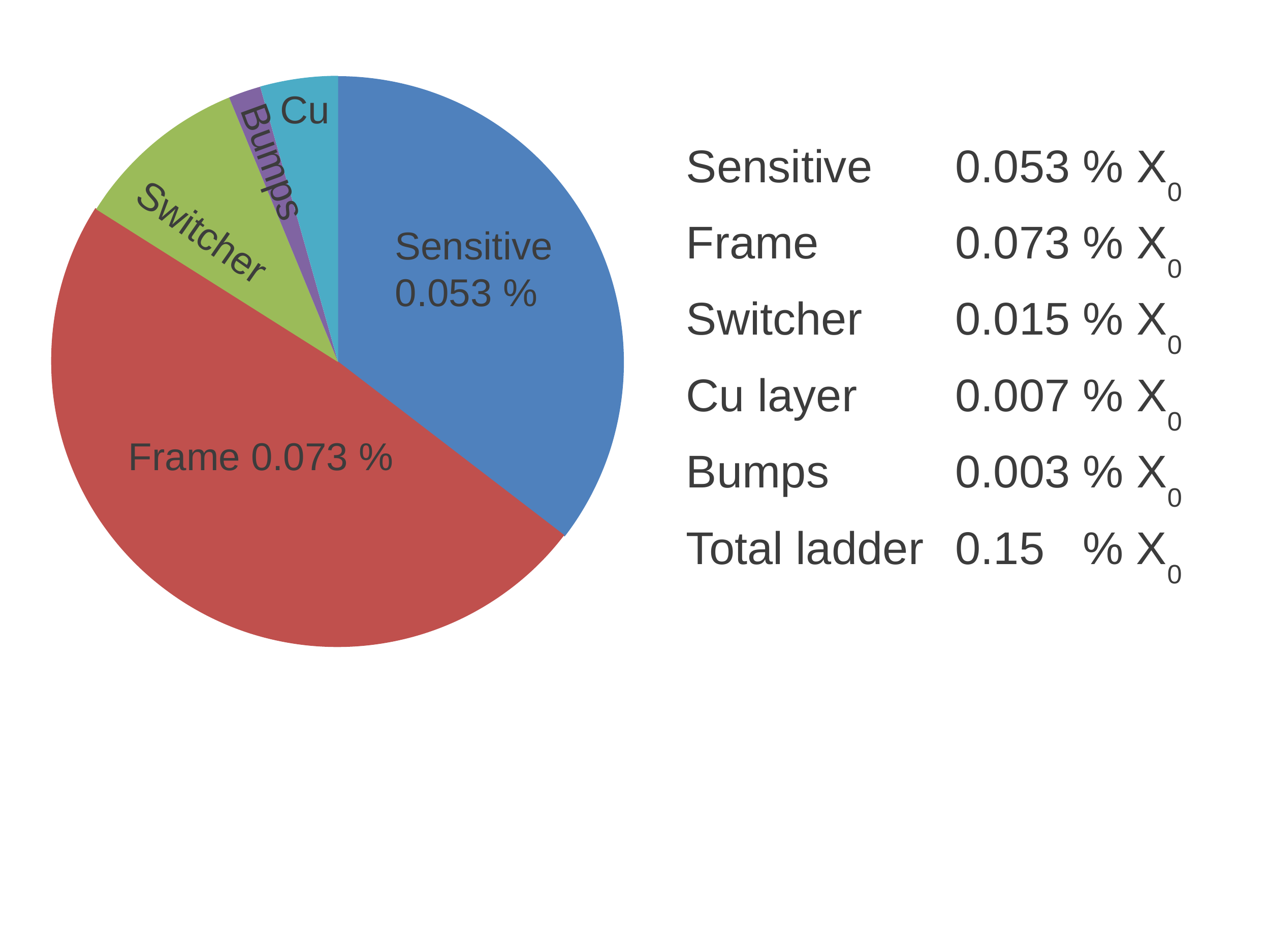}
\vskip -2cm
\caption{Break-down among the detector components of the material expressed in units in radiation lengths over the acceptance of the all-Silicon ladder design. }
\label{fig:material}
\end{figure}

The detailed ladder design for an LC vertex detector envisages a contribution 
to the detector material (averaged over the ladder area) that is equivalent 
to $\sim 0.15$\% X$_0$/layer. A break-down of the material among the 
components with the most important contributions is presented in 
Figure~\ref{fig:material}. The fully engineered Belle~II design~\cite{b2tdr} 
corresponds to $0.21$\% X$_0$/layer over the active area. The 
LC budget is reduced primarily by a more aggressive thinning of
the sensor. The thickness of the active material, one of the dominant
contributions, is reduced from 75 $\mathrm{\mu m}$ to 50 $\mathrm{\mu m}$ 
(as already achieved in the Belle~II prototype production run). 
The support frame is reduced from 450 $\mathrm{\mu m}$ to 400 $\mathrm{\mu m}$.

\section{Sensor production}
\label{sec:sensors}

The DEPFET pixel technology contains 25 photolithographic mask steps and nine
implantations. The process involves moreover very different technology 
aggregates like wafer bonding, Silicon-on-Insulator (SOI), double poly 
Silicon, triple metal (including copper), back side thinning and double 
sided wafer processing. 
All processing steps except the SOI production
are performed in-house. 

Two early sensor production batches have proven the feasibility 
of small pixels down to 20 $\times$ 20 $\mathrm{\mu m^2}$. The performance of 
the ILC prototypes constructed with these sensors is described in several 
previous publications~\cite{Andricek:2009zz, Velthuis:2008zza}.

The more recent Belle~II prototype production run (``PXD6'') was performed 
on an SOI wafer and all matrices were thinned to 50 $\mu \mathrm{m}$. 
Sensor designs with large pixels 
(50 $\times$ 50 $\mu \mathrm{m}^2$ and 50 $\times$ 75 $\mu \mathrm{m}^2$) 
for Belle~II were added to the mask, 
as well as further design variations of the pixel layout. The results of 
a characterization of these sensors are presented in Section~\ref{sec:tb}.

Based on the results and experiences of ``PXD6'' prototyping, the production 
run for Belle~II (``PXD9'') has started. Apart from some other, minor 
modifications, the gate and clear-gate dielectrics are reduced to 100 nm, 
approximately half the thickness used previously, for better radiation 
hardness\footnote{The Belle~II
environment, with an expected ionizing dose of 
18.5 $\mathrm{kGy}/\mathrm{yr}$, 
exposes the sensors to significantly higher radiation 
levels than that at the ILC, where the required tolerance amounts 
to approximately 1 kGy/year~\cite{ilddbd}. }. This leads to a reduction
of the internal gain, that scales approximately with the square root of
the oxide thickness. The performance is partially recovered by modifying 
other parameters of the FET layout. The gain depends strongly on the 
gate length ($g_q \propto L^{-3/2}$) and in the longer term a considerable 
increase of the improvement can be achieved by reducing the gate dimension. 
A more detailed discussion of the dependence of the gain is 
left for Section~\ref{sec:tb}).

\section{Control and read-out ASICs}
\label{sec:asics}

Several auxiliary ASICs are required to operate the DEPFET sensor. The most important developments
are briefly summarized in this Section.

The SWITCHER control chips select segments of the sensor (pairs of rows or
4-row segments) for read-out. A separate driver supplies the {\em clear} 
pulse of up to 20 V to remove the collected signal from the internal gate 
after read-out.

\begin{figure}[h!]
 \centering
 \includegraphics[width=\columnwidth]{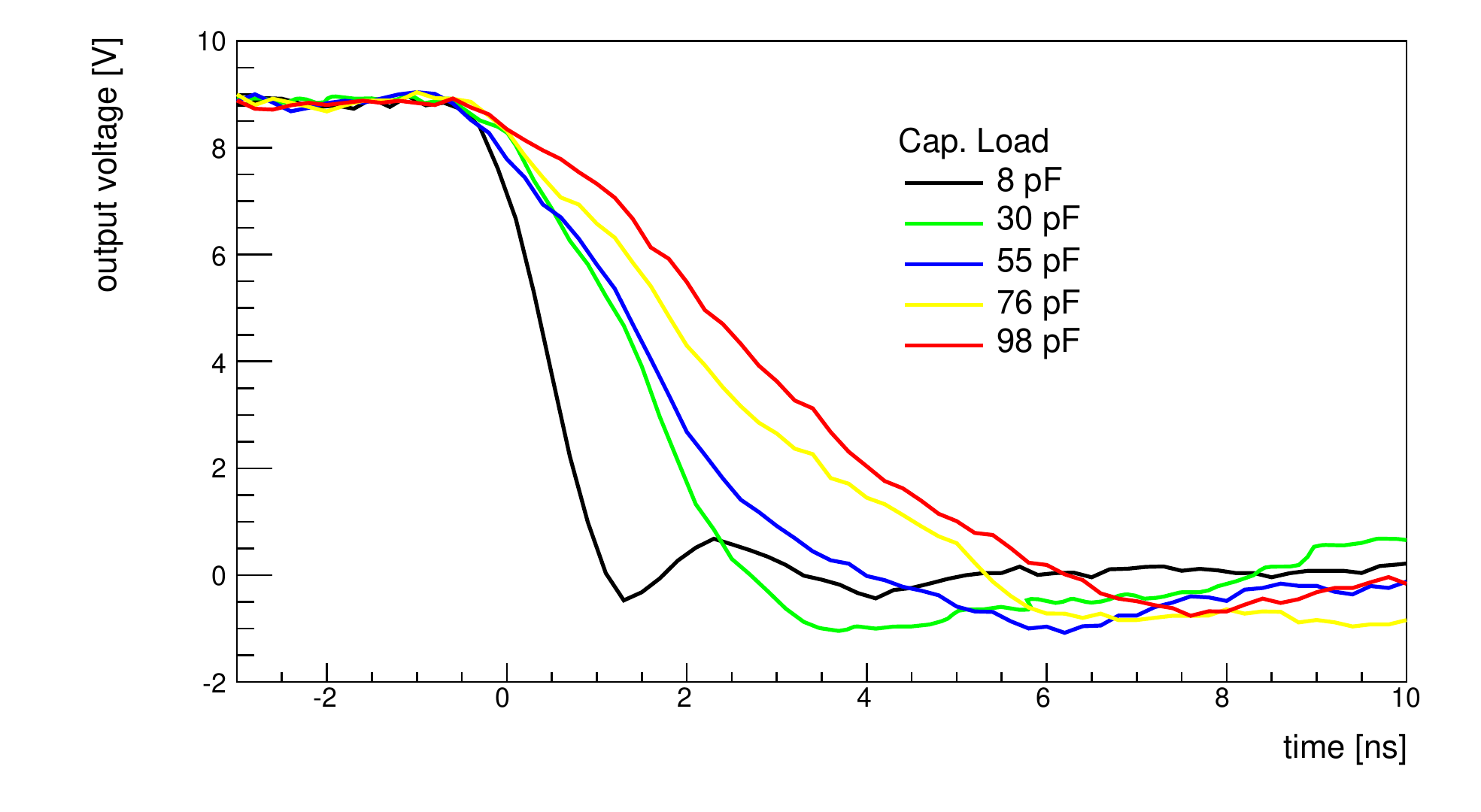}
  \caption{The falling edge of a 9 V pulse produced by SWITCHERB18, measured with a number of different load capacitances.}
\label{fig:switcher}
\end{figure}

Two designs of the SWITCHER versions optimized for Belle~II 
requirements (SWITCHERB18 in 0.18 $\mu m $ and SWITCHERB in 0.35 $\mu m$) have 
been produced and tested successfully. 
The radiation hardness of the 0.35 $ \mu m$ design is demonstrated 
up to 370 kGy. The 0.18 $ \mu m$ version is smaller, allows 
for faster switching, lower power consumption and is expected to improve the 
radiation tolerance. It can produce pulses with a maximum swing of 20 V
in a voltage range of 50 V. The falling edges under different load conditions 
of 9 V pulses generated by SWITCHERB18 are presented in 
Figure~\ref{fig:switcher}. These measurements show that the current
SWITCHER chips can drive long lines, up to loads well beyond the requirements 
of the Belle~II and LC vertex detectors.

The drain current signals from 256 columns of pixels are processed and 
digitized by the DCD (Drain Current 
Digitizer~\cite{Knopf:2011zz, Peric:2010zza,asi:papdcd2, thesisjochen}) chip. 
The analog input stage keeps the column line 
potential constant (necessary to achieve fast readout), compensates for 
variations in the DEPFET pedestal currents, and amplifies and shapes the 
signal. The current receivers in the most recent DCD versions are based 
on transimpedance 
amplifiers (replacing the regulated cascodes of earlier generations).
The analog signal is digitized using two 8-bit current-mode cyclic ADCs
with a sampling frequency of 10 MSamples/s. The DCD can be operated in
double correlated sampling mode or single sampling mode. The latter is
preferred as it allows higher read-out speed.

The DCD is implemented in UMC 0.18 $\mu m$ CMOS technology using special 
radiation hard design techniques (e.g. enclosed NMOS gates) in the analog part.
The 256-channel DCDB, with an area of 3.2 $\times$ 5 $\mathrm{mm}^2$, is 
optimized specifically for Belle~II requirements. 
The DCDB chip is fully functional at the nominal 320 MHz. Radiation 
tolerance of at least 70 kGy has been proven.

The derandomized raw data from the DCD are transmitted to the Data Handling 
Processors (DHP~\cite{Lemarenko:2012zz}) using fast parallel 8-bit digital 
outputs. This third ASIC, 
located on the end-of-ladder area immediately behind the DCD, performs data 
processing (pedestal subtraction, common code correction), compression
(zero suppression), buffering and fast serialization. It furthermore 
controls the other read-out ASICs.

The first full-scale DHP prototype was implemented in IBM 90 nm 
CMOS technology. Communication of the DHP chip with the other ASICs 
has been successfully tested. Implementation of
the next DHP version is foreseen in TSMC 65 nm CMOS technology. 
In the longer term the DCD and DHP are envisaged to evolve into a single chip.

\begin{figure}[h!]
 \centering
 \includegraphics[width=\columnwidth]{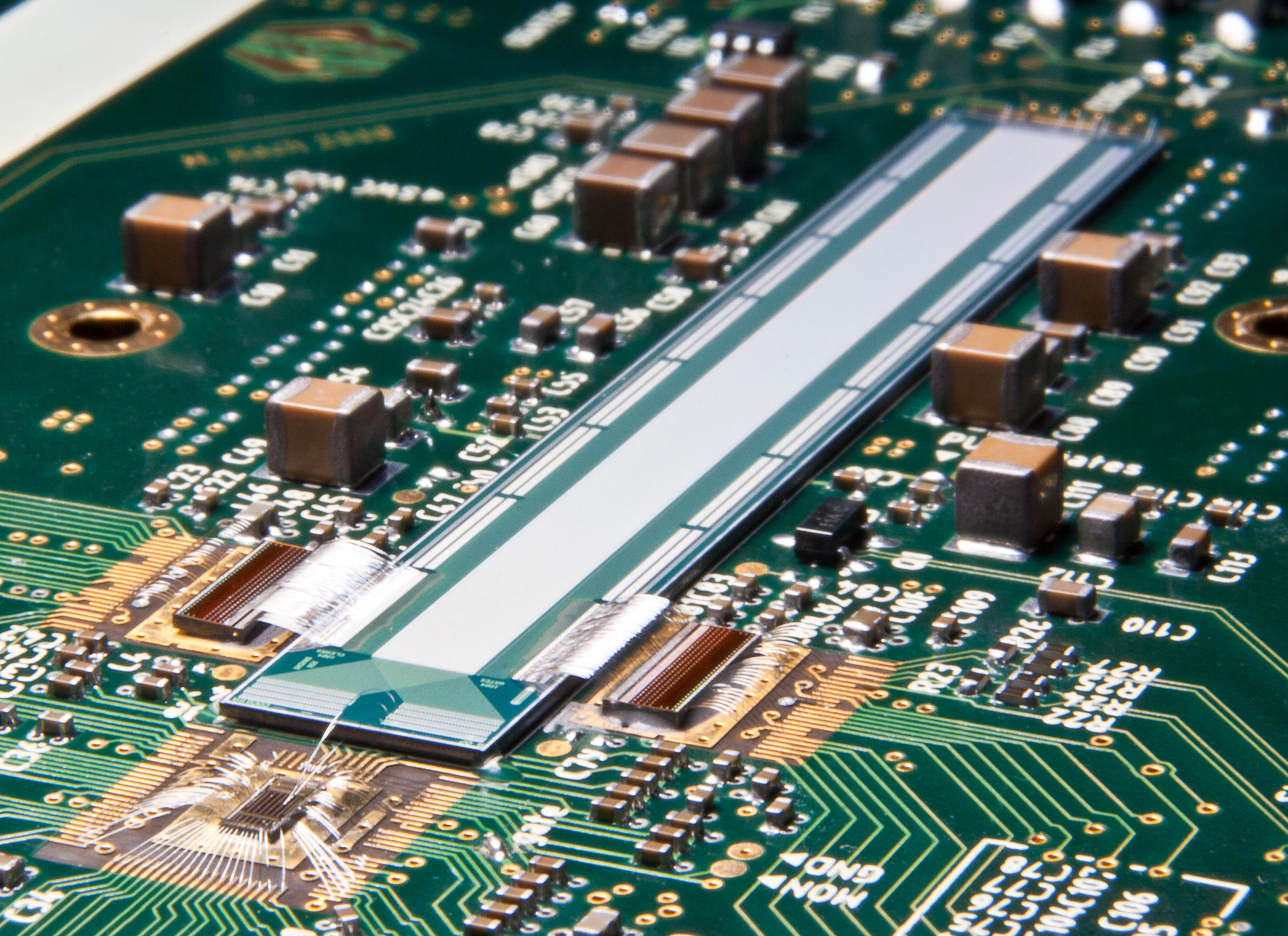}
  \caption{DCD2 (the precursor of DCDBv2), SWITCHER and a DEPFET ladder mounted on a read-out board. Note that in this early test system only a small number of columns are connected through wire bonds and only one SWITCHER was connected.}
\label{fig:testboard}
\end{figure}

The analog stage of the DCD is most important for the detector performance. 
The analog response has been characterized
in detail~\cite{thesisjochen, thesismanuel} on the test system of 
Figure~\ref{fig:testboard} where a DCD2 and SWITCHER are connected 
to a full-size DEPFET sensor.

\begin{figure}[h!]
 \centering 
  \includegraphics[width=\columnwidth]{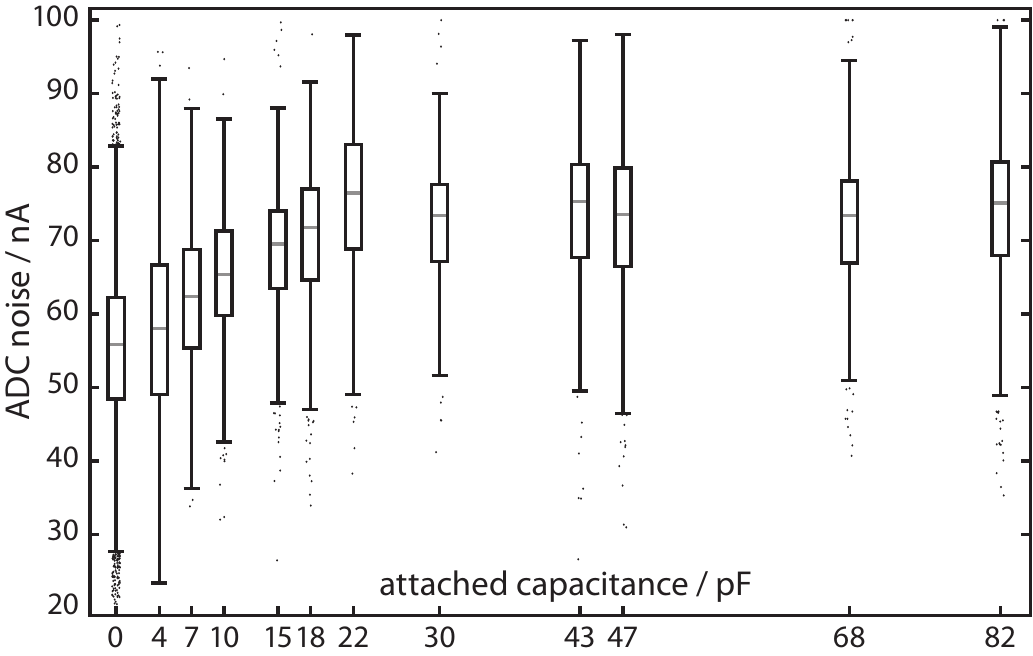} \\
\vskip 2mm
 \includegraphics[width=\columnwidth, page=2]{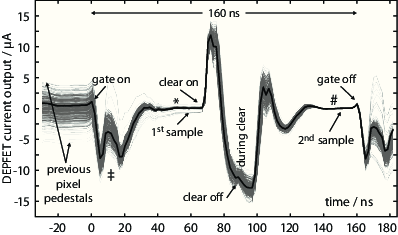} 
 \caption{DCD2 noise versus capacitative load on the input stage of the ASIC 
(upper panel). The load of a full-length DEPFET half-ladder corresponds to 
approximately 80 $\mathrm{pF}$. Time-resolved measurement of the DEPFET drain 
current during the read-out sequence (lower panel). 
The most important events in the read-out cycle are indicated on the figure. 
}
\label{fig:measurementsmanuel}
\end{figure}

The ADC conversion factor is found to be approximately 
10.2 $\mathrm{LSB} / \mu \mathrm{A}$, with a slight
dependence on the read-out speed and the analog supply voltage, with
a channel-to-channel variation of less than 
0.2 $LSB / \mu \mathrm{A}$~\cite{thesismanuel}. 
In the upper panel of Figure~\ref{fig:measurementsmanuel} noise measurements 
for several values of the capacitative load on the input stage are presented. 
We find a DCD noise of approximately 80 $n \mathrm{A}$ for the load of a 
full-length DEPFET ladder (about 80 $\mathrm{pF}$).

The read-out speed is one of the most challenging requirements of the LC vertex
detector. The sequence involved in the read-out of one row in single-sampling 
mode is illustrated in the lower panel of Figure~\ref{fig:measurementsmanuel}. 
The drain current 
is found to settle approximately 30 ns after the row is activated (indicated 
as ``gate on'' in the figure). After the current is sampled the clear pulse 
is applied. Thus a row rate of approximately $\mathrm{1/100}$ ns is achieved. 
Figure~\ref{fig:measurementsmanuel} clearly shows that there is ample room 
for improvement. A further increase of the read-out speed, with row rates
down to $\mathrm{1/40}$ ns, requires a number of changes to the design of 
the read-out ASIC.

To interprete this result in the LC perspective we consider the following 
layout: the 12.5 cm long inner layer of the ILC VXD is equipped with
a DEPFET sensor with read-out ASICs on both ends. Pixels in the center of the
sensor are 25 $\times$ 25 $\mathrm{\mu m^2}$. The pixel size is varied 
over the length of the sensor, ensuring that charge is shared over
a small number of pixels independent of the z-position. The pitch is 
increased to 50 $\mathrm{\mu m}$ 
at $|z| = \pm$ 1 cm and to 100 $\mathrm{\mu m}$ at $|z|= \pm$ 2 cm. 
With a column depth of 1025 pixels per half-ladder, and two (four) rows 
sampled in parallel, the envisaged row rate 
of $\mathrm{1/80}$ ns implies a read-out time for a complete frame of 
40 $\mu \mathrm{s}$ (20 $\mu \mathrm{s}$).




\section{Power consumption \& Cooling concept}
\label{sec:cooling}

The strict material budget of the LC vertex detector forces a tight control
of the power consumption. The DEPFET sensor with rolling shutter read-out 
are intrinsically well suited to this environment, as its power consumption 
scales with the number of pixels addressed in parallel (i.e. the number of 
read-out columns), rather than with the number of pixels. 
In the DEPFET sensor, the signal is collected in the internal gate 
regardless of the gate voltage supplied to the FET. The FETs only need to be 
powered during the brief lapse of time when the drain current is read out.
Only a small fraction of rows is therefore
active at any time during operation. All other pixels are switched off and draw 
no current. Other sources, such as the leakage current are negligible 
even after irradiation.

For the layout sketched in Section~\ref{sec:asics}, and assuming 2-fold 
(4-fold) read-out only 0.2\% (0.4\%) of pixels are active at any time 
during operation. 
The instantaneous power consumption of a half ladder is therefore given by: 
\begin{flalign*}
\begin{array}{@{}>{\displaystyle}l@{}>{\displaystyle{}}l@{}}
\hskip 1cm P &= I_d \times V_{ds} \times n_{col} \times n_{\|} \\
\hskip 1cm  &\sim 100\,\mu \mathrm{A} \times 5\,\mathrm{V} \times 1000 \times 2 = 1 \mathrm{W}, 
\end{array}
&&
\label{equation:P_diss}
\end{flalign*}
where $I_d$ is the drain current, $V_{ds}$ is the source-drain voltage, 
$n_{col}$ the number of columns in the sensor and $n_{\|}$ the number of 
rows read out in parallel. The estimate given corresponds to the situation
where two rows are read out in parallel. For four-row read-out the power
consumption in the sensor is doubled. 


The power consumption required to drive the gate and clear lines is dominated 
by the SWITCHERs that consume 225 $\mathrm{mW}$ per active row. The consumption
of each active chip is 50 mW and idle chips contribute 10 $\mathrm{mW}$ each. 
With two active rows per half ladder at any time during operation the 
power consumption is approximately 500 $\mathrm{mW}$ per half ladder. 

The current consumption of the 256-channel DCD is measured to be
1.3 W. Most power is dissipated in the analog input stage that
requires up to $\sim$ 3 $\mathrm{mW}/$channel to ensure fast read-out with
good noise performance. The DHP implemented in a 90 nm process is found
to consume less than 200 mW. Four DCDs and four DHPs are required to 
read out a half-ladder, leading to a local power consumption at each end 
of the ladder of 6~W.

With 8 short and narrow inner ladders in layer 1 and 56 wider and longer outer 
ladders in layers 2 through 5, the total instantaneous power required
to operate a DEPFET vertex detector at the ILC is approximately 1~kW.
Note that these results, based on the measured consumption of complete
ASICs, are slightly better than the conservative estimate
in 2007~\cite{VXDreview}. 

The LC duty cycle of 1/275 is the key to reduce the average power consumption
of the detector further. The power pulsing scheme pursued by most LC detectors
is implemented in DEPFET sensors for XFEL, where the DEPFET and the analogue
part of the electronics are switched off between bursts. Taking an effective
detector duty cycle of 1/100 the average power consumption of the vertex
detector can be kept below 10~W.

The DEPFET cooling concept was developed on the basis of a detailed
finite element model~\cite{VTXdepfet:mechmarinas} of the read-out module.
For crucial parameters the simulation is cross-checked against measurements 
on test structures. As the Belle~II vertex detector relies on
a forced flow of dry and cold gas to reduce the temperature
of the center of the ladder, the potential of gas cooling was studied
with special care. To predict the temperature in the LC the same
model is adapted. 

\begin{figure}[h!]
\centering
\includegraphics[width=\linewidth]{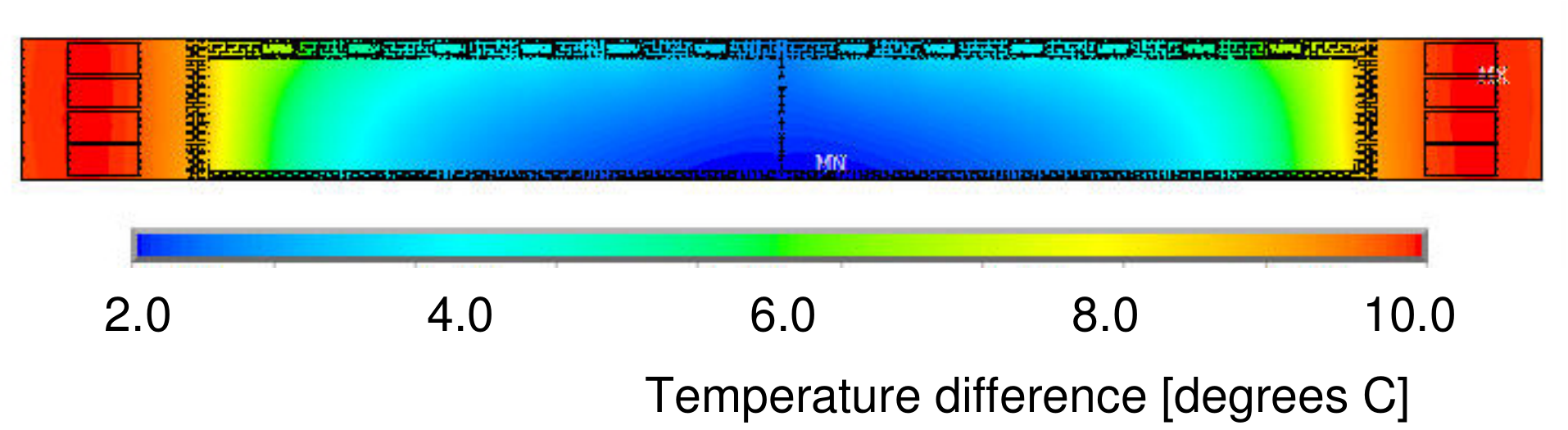}
\caption{Temperature distribution along a DEFPET ladder referenced to the gas temperature. The power consumption of the different detector components is as described in the text. A factor $1/25$ is applied to take into account the power pulsing. The ladder is cooled by a forced flow of 2 $\mathrm{m/s}$ of cold gas. }
\label{fig:thermalsimulation}
\end{figure}

The power consumption of the different detector components
is as described above, leading to a total instantaneous power consumption
of 600~W. The average power is taken as $1/25$ of the 
instantaneous consumption to account for the impact of
pulsed powering. A gas flow of 2 $\mathrm{m/s}$ is forced
on both sides of the ladder. The expected temperature distribution 
for the innermost LC ladder 
is shown in Figure~\ref{fig:thermalsimulation}. The end-of-ladder
with the read-out electronics reaches a temperature 10$^\circ$ C 
above that of the cold gas, while the center of the sensor stays
within two degrees of the gas temperature.

\begin{figure}[h!]
 \centering 
 \includegraphics[width=\columnwidth]{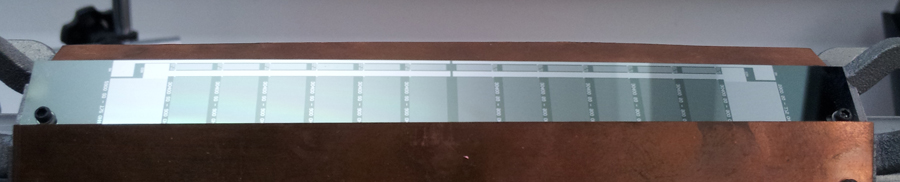}
 \caption{Photograph of a thermo-mechanical sample. The circuits that mimic the ASICs are clearly visible.}
\label{fig:thermalimage}
\end{figure}

The understanding of the thermal properties of the ladder is greatly 
enhanced by measurements on the thermo-mechanical samples shown in
Figure~\ref{fig:thermalimage}. 
These are mechanically identical to DEPFET all-Silicon ladders, 
but lack some of the processing steps needed to turn them
into fully functional sensors. Small circuits on the relevant positions 
allow to mimic the power consumption of the ASICs and the sensor.
The power dissipated in each circuit can be regulated independently.

Using thermo-mechanical samples a mock-up has been built that represents
the first two layers of the vertex 
detector~\cite{VTXdepfetmechair, thesispablo}.
The mock-up is equipped with a system to provide a gas flow of controlled 
rate, temperature and humidity. Measurements of the temperature
distribution with a thermal camera and an environmental monitoring system 
based on Bragg fibers~\cite{Moya:2012zh} confirm the predictions of the 
finite element model. 

\begin{figure}[h!]
 \centering 
  \includegraphics[width=\columnwidth]{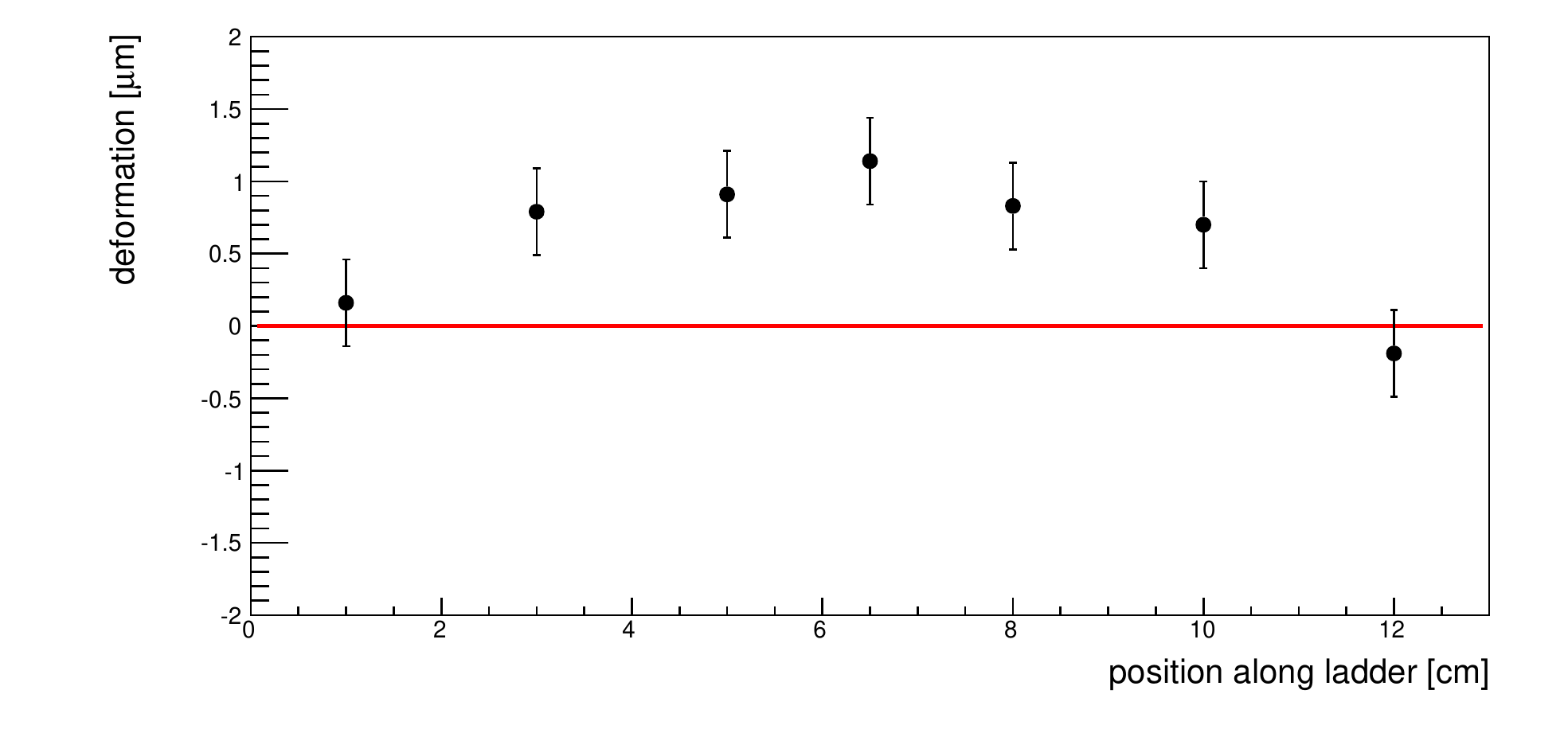}
 \caption{The displacement perpendicular to the ladder due to the application of a forced gas flow versus position along the ladder. The gas pressure is 4 bar, well over the nominal pressure required to cool the sensor.}
\label{fig:vibrations}
\end{figure}

Capacitative sensors are used to monitor the
position of the ladder without distorting the system. Upon application of the
gas flow the ladder position is slightly distorted. The maximum deformation
is registered at the center of the sensor: the measured $r$-coordinate 
changes by 1.1 $\pm$ 0.3 $\mu \mathrm{m}$ when the gas flow is switched on. 
No significant hysteresis is observed; the original ladder position is
recovered when the gas flow is switched off.
Vibrations introduced by the gas flow are detected by performing 
position measurements with and without gas flow and taking the difference
of the Fast Fourier Transforms of both time series. 
A clear peak at 400 $Hz$ is observed with an amplitude of 1.2 $\mu \mathrm{m}$ 
(0.7 $\mu \mathrm{m}$) for a gas pressure of 10 bar (4 bar). 
The magnitude of the static and dynamic deformations of 
1-2 $\mu \mathrm{m}$ perpendicular to the ladder have a negligible impact 
on the resolution of the vertex detector.

\section{Characterization of prototypes}
\label{sec:tb}

The response of DEPFET sensors from the PXD5 and PXD6 production has been
characterized in beams of charged particles from accelerators at CERN and 
DESY~\cite{Andricek:2009zz,Velthuis:2008zza,Reuen:2006zf,Kohrs:2006kk}. 
In the following only some highlights from this extensive program are 
presented.

The first set of measurements is on 450 $\mathrm{\mu m}$ ILC-design sensors 
with pixel sizes ranging from 20 $\times$ 20 $ \mathrm{\mu m^2}$ to 
24 $ \times $ 32 $ \mathrm{\mu m^2}$. The read-out module for these 
prototypes relies on the CURO chip~\cite{Kohrs:2005ir,thesistrimpl}. 
More recent measurements 
correspond to thinned PXD6 sensors with DCDB read-out.

\begin{figure}[h!]
 \centering 
  \includegraphics[width=\columnwidth]{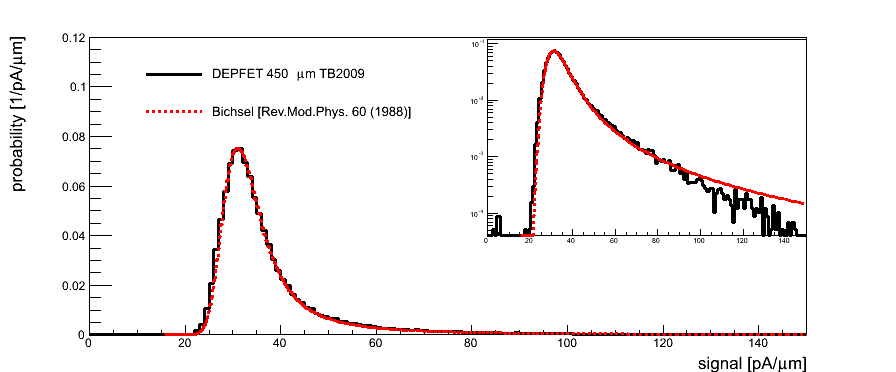}
 \includegraphics[width=\columnwidth]{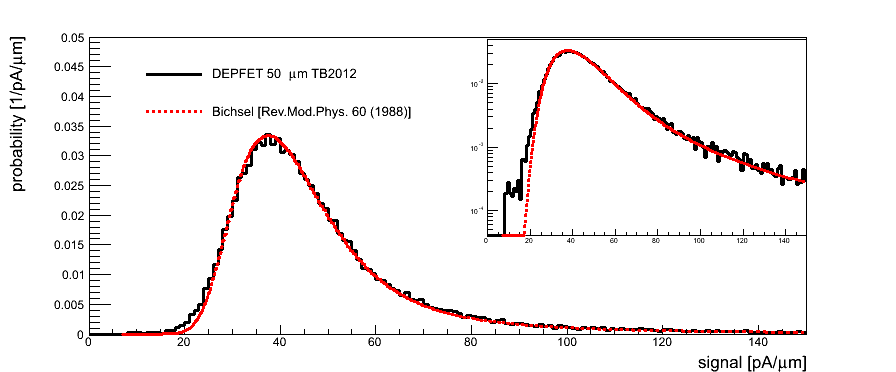}
 \caption{Signal distribution for a 450 $\mathrm{\mu m}$ thick DEPFET 
sensor (upper panel) and for sensor thinned to 50 $\mathrm{\mu m}$ 
(lower panel). In both cases the measurement is expressed in pA per micron 
of active Silicon. The measurements are compared to the prediction for the 
energy deposition in thin Silicon layers of H. Bichsel~\cite{bichsel}. The 
deposited energy is converted to signal per unit thickness by dividing by the 
ionization potential of Silicon (3.62 eV) and multiplying by the quantum gain 
$g_q$ of the sensors (340 $\mathrm{pA}/e^-$ for the 450 $ \mathrm{\mu m} $ 
thick sensor, in good agreement with the result of a calibration with a 
$\gamma$ source, and 470 $\mathrm{pA}/e^-$ for the thin sensor).}
\label{fig:landau}
\end{figure}

The DEPFET drain current distributions due to perpendicularly incident 
120~GeV pions from the CERN SPS are shown in Figure~\ref{fig:landau}.
The upper panel corresponds to a 450 $\mathrm{\mu m}$ thick ``PXD5'' sensor 
and the lower panel to a PXD6 sensor thinned to 50 $ \mathrm{\mu m}$. 
The signal in adjacent pixels is added using a simple clustering 
algorithm with a neighbor cut of 2.6 $\sigma$. Further details of 
the analysis can be found in References~\cite{benjamin,thesismarca}.
Both measurements are corrected for the gain of the Front End 
ASICs\footnote{The gain measurement for the DCDB is presented in 
Section~\ref{sec:asics}. For the CURO system used to read out the thick 
sensor a conversion factor of 7.7 nA/LSB is used~\cite{thesisrobert}.}. 
To facilitate comparison the results are moreover divided by the sensor 
thickness.
 
The observed distributions are in good agreement with the model prediction
of H. Bichsel~\cite{bichsel}, where only the internal gain of the sensors
is left floating. In particular, the model correctly predicts
the broader signal distribution observed for the thin sensor: The peak position 
divided by the Full Width at Half Maximum (FWHM) yields: $ \Delta_p / w = $
1.67 for the 50 $ \mathrm{\mu m}$ sensor (prediction: 1.61) and 
$ \Delta_p  / w = $ 3.06 for the 450 $ \mathrm{\mu m}$ thick sensor 
(prediction: 3.13). 

\begin{figure}[h!]
 \centering 
  \includegraphics[width=\columnwidth]{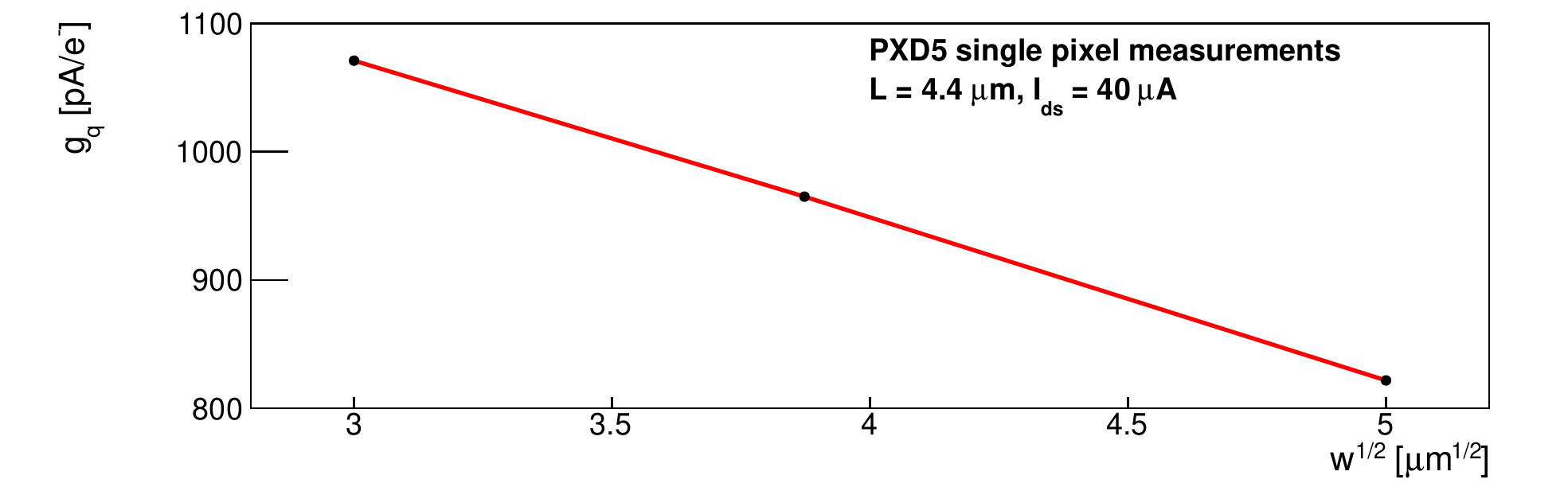} 
  \includegraphics[width=\columnwidth]{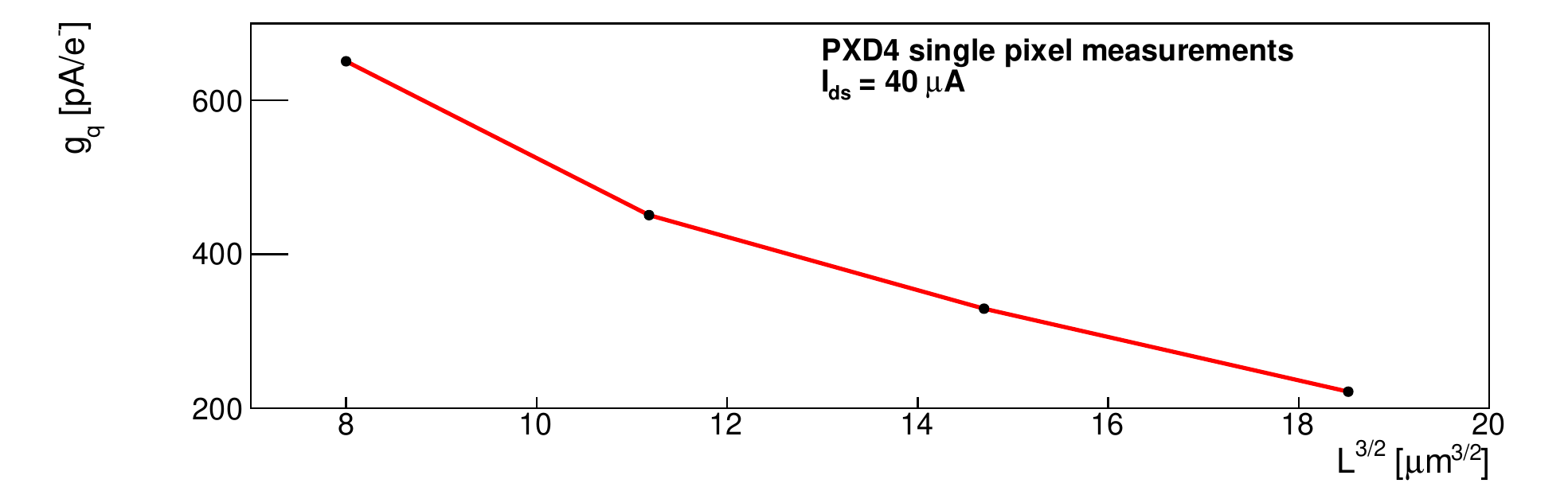} 
  \includegraphics[width=\columnwidth]{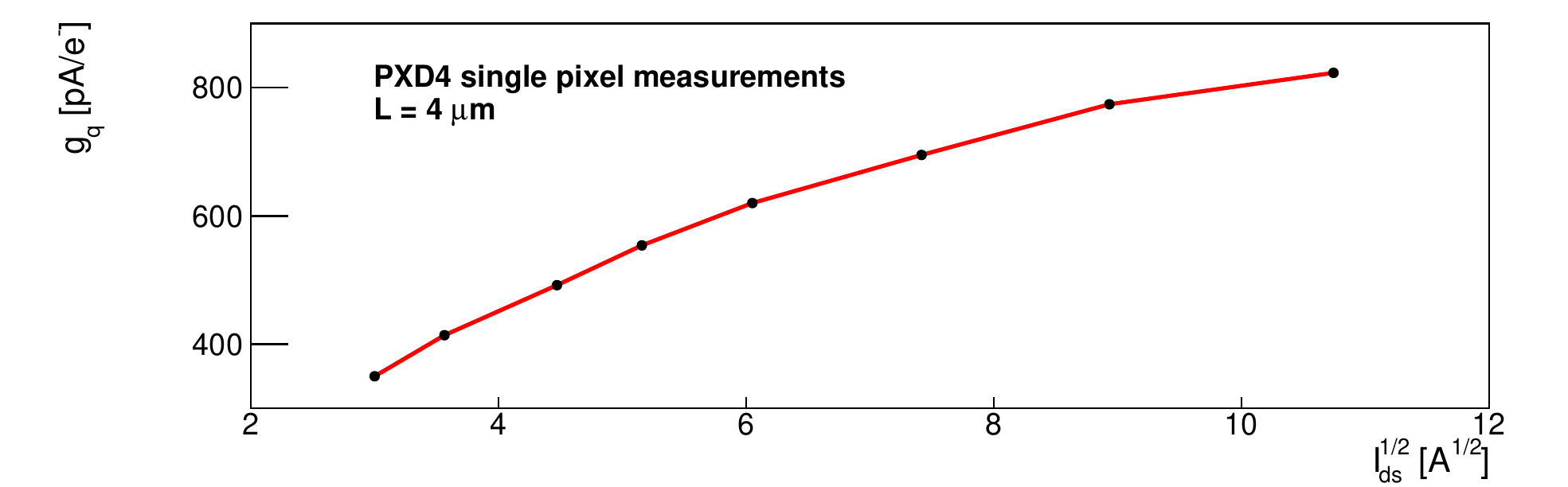}
 \caption{Gain measurements on single-pixel structures. The axes are chosen such that, according to Equation~\ref{eq:gain}, the measurements should lie on a straight line.} 
\label{fig:gqdependence}
\end{figure}

\begin{figure}[h!]
 \centering 
  \includegraphics[width=\columnwidth]{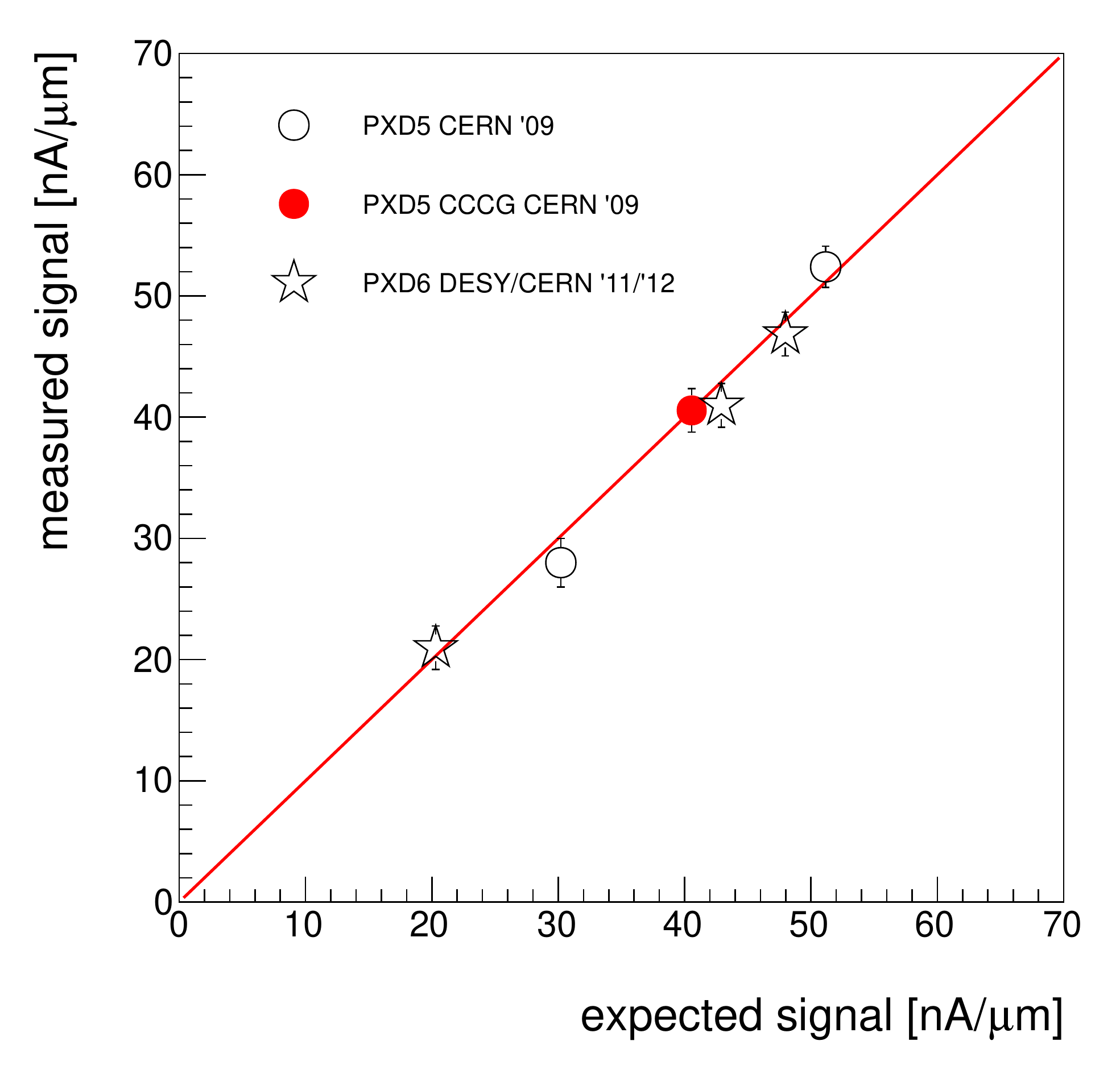}
 \caption{Overview of the measurements of the internal gain of DEPFET prototype sensors. The expected signal is obtained by scaling the response of the CCCG module using the simple module discussed in the text. The expectation for the PXD5 modules without CCCG is moreover scaled by 0.75 to account for the incomplete clear in these devices. }
\label{fig:gq}
\end{figure}

The gain of the in-pixel amplification stage depends on the layout of the FET,
 in particular on the length of the gate $L$, its width $w$ and the oxide
thickness $t_{ox}$, and on the drain-source current $I_{ds}$. A simple 2D
model predicts the following relation:
\begin{equation} 
g_q \propto \frac{I^{1/2}_{ds} \times t^{1/2}_{ox}}{w^{1/2} \times L^{3/2}}
\label{eq:gain}
\end{equation}
Examples of single-pixel gain measurements from 
Reference~\cite{Andricek:2012zz, thesisrummel} are shown in 
Figure~\ref{fig:gqdependence}. The response of the complete matrices 
is compared to the prediction of the model in Figure~\ref{fig:gq}. 
The $g_q$ measurements on a large number of small-scale 
structures~\cite{Andricek:2012zz, thesisrummel} are also drawn.
Clearly, the simple 2D model is insufficient to correctly predict
the dependence. In particular, the assumption of constant carrier 
mobility is known to fail for very large drain current. With some caution
as to the range of applicability the model provides, however, an adequate set of
rules of thumbs.

The uniformity of the signal over the area of the sensor has been evaluated
as well. In a PXD5 device with approx. 8000 pixels and a 4 
$ \mu \mathrm{m} $ gate 
length a gain spread of 5\% $\pm$ 3\% is observed. A more recent device with a 
5 $\mu \mathrm{m}$ gate length shows an even more uniform
response (random gain spread $<$ 3\%), but has a systematic difference of 
5\% between even and odd rows of the device. The cause of this effect
is currently being investigated. Variations at this level have a negligible 
impact on the overall detector performance.


Charge sharing between adjacent pixels is well understood. For typical
operating parameters and perpendicularly incident particles the signal 
in the thin sensors is contained in an 
area with a diameter of several tens of microns. In sensors with large
pixels (50$-$75 $ \mu\mathrm{m}$) an extra drift ring is inserted into 
each pixel to ensure rapid and efficient charge collection.

\begin{figure}[h!]
 \centering 
  \includegraphics[width=\columnwidth]{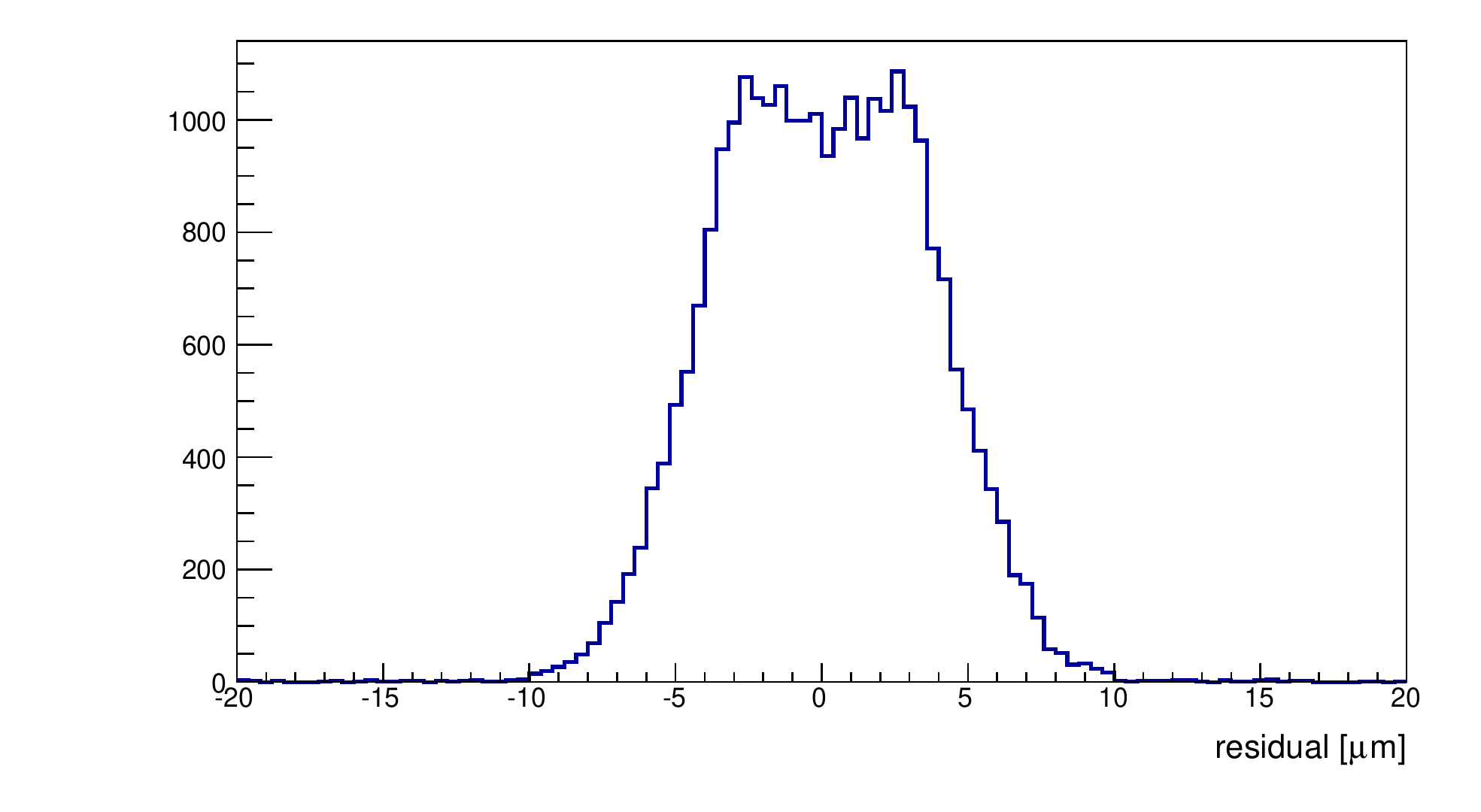}
 \caption{Simulated residual distribution (reconstructed $-$ true position) for an LC design DEPFET sensor (20 $\times$ 20 $\mu \mathrm{m}^2$ pixels on a 50 $\mu \mathrm{m}$ thick sensor). The simulation mimics the beam test conditions with high energy pions traversing the sensor under a 90 degree incidence angle. The RMS of the distribution is 3.5 $\mu \mathrm{m}$.}
\label{fig:residuals}
\end{figure}

The spatial resolution of DEPFET devices has been studied in detail. 
Thick DEPFET sensors with an ILC pixel design 
(thickness $ \times $ pixel size of 450 $\times$ 20 $\times$ 20 $\mu m^3$) 
approach a 1 $ \mu m$ resolution in both 
coordinates~\cite{Andricek:2011zza,Andricek:2009zz,Velthuis:2008zza}. 
The thin Belle~II design
sensors with 50 $ \times $ 50 $\times$ 75 $\mu \mathrm{m}^3$ pixels 
yield a spatial resolution of 8 $\mu \mathrm{m}$~\cite{thesismarca} for
perpendicularly incident tracks, in good agreement with the expected
performance. Using an extensive set of beam test results on both
types of devices, a detailed model of the DEPFET response has been developed 
for use in simulations of the overall detector performance. This {\em digitizer}
model is used to predict the spatial resolution of an LC-design DEPFET sensor
with a detector thickness of 50 $\mu \mathrm{m}$. For comparison to beam 
test results we consider high-energy charged pions traversing the sensor
under a 90 degree incidence angle in the residual distribution of 
Figure~\ref{fig:residuals}. For a DEPFET device with
20 $\times$ 20 $\mu \mathrm{m}^2$ (25 $\times$ 25 $\mu \mathrm{m}^2$)
pixels on a 50 
$\mu\mathrm{m}$ thick sensor a spatial resolution of 3.5 (4.5) 
$\mu \mathrm{m}$ is found. In more realistic conditions the performance 
is expected to be significantly better than that. Most tracks traverse the 
sensor under an angle and deposit signal
in multiple pixels. The large magnetic field moreover causes the charge 
carriers to drift to the read-out plane under an angle (the Lorentz angle). 
Charge sharing between neighbouring pixels leads to a significant improvement
in the spatial resolution. 

\section{Summary}
\label{sec:conclusions}

Future colliders require vertex detectors that combine an excellent 
spatial resolution and challenging read-out speed with a very tight 
material budget. The Depleted Field Effect Transistor (DEPFET) offers a 
promising way to address these challenges. A matrix of DEPFETs 
embedded in a detector-grade 
Silicon sensor forms an active pixel sensor with competitive characteristics.
A novel thinning concept~\cite{Moser:2007zza,Andricek:2004cj}, 
known as the all-Silicon ladder, allows to 
produce self-supporting ladders with a minimal material budget 
($\sim $ 0.15\% $X_0$/ladder averaged over the ladder area 
for an LC vertex detector).

The DEPFET collaboration has succesfully produced
small-pixel sensors (down to 20 $\times$ 20 $\mathrm{\mu m^2}$) and
thinned sensors to 50 $\mathrm{\mu m}$ thickness. The design
of the Belle~II sensors is frozen and the production of sensors
for that experiment has been started. The control
and read-out ASICs required to operate the sensor have been
produced and are found to meet the LC specifications for
read-out speed and noise performance.

The power consumption in the DEPFET sensors with rolling shutter read-out 
scales with the number of pixels read out in parallel. An estimate for a
DEPFET vertex detector at the LC, based on the measured 
consumption of the current generation of read-out and control ASICs, 
yields an instantaneous power consumption of a~kW. Assuming
a duty cycle of 1/100 the average consumption corresponds
to only 10~W. Finite element simulations and measurements on a mock-up
equipped with thermo-mechanical dummy sensors show that a forced
flow of cold gas is very effective in removing the heat from the
system, while causing negligible deformations of the 
ladder.

Prototypes based on thin Belle~II design sensors equipped
with the ASICs to be used in the experiment have been subjected
to an extensive test program. The response to minimum ionizing
particles matches closely with the expectation.
The dependence of the in-pixel gain on the FET design parameters 
(gate dimensions and oxide thickness) is found to agree approximately 
with a naive 2D model. These result show that extremely thin
DEPFET ladders (down to 50 $\mu \mathrm{m}$) can achieve a
comfortable S/N ratio of over 40. A detailed model 
of the DEPFET response that provides and adequate description of 
the signal cluster properties and spatial resolution of the prototypes 
submitted to beam tests, are used to predict a spatial resolution of
3.5 $-$ 4 $\mu \mathrm{m}$ for perpendicularly incident MIPs on an
LC design DEPFET device.

We believe, therefore, that the DEPFET sensor technology can fulfill 
the requirements for a pixel detector in a future LC experiment.

\section*{Acknowledgments}

The authors would like to acknowledge the support from the AIDA project, 
that provided the infrastructure used to characterize the DEPFET prototypes
in beam tests at CERN and DESY. We also thank Theresa Obermann for her help.

We furthermore thank Hans Bichsel for access to the predictions of his model
and him and the authors of GEANT4 
for the interesting discussion on the expected signal.

This work was supported by the German Ministerium f\"ur Bildung, Wissenschaft, 
Forschung und Technologie (BMBF), by the Ministry of Education, 
Youth and Sports of the Czech Republic and by the Spanish Ministerio 
de Ciencia e Inovaci\'on. Marcel Vos is supported by the Spanish 
Ram\'on y Cajal program.

\vfill

\begin{IEEEbiography}[{\vskip -1cm \includegraphics[height=1.5in,width=1in,clip,keepaspectratio]{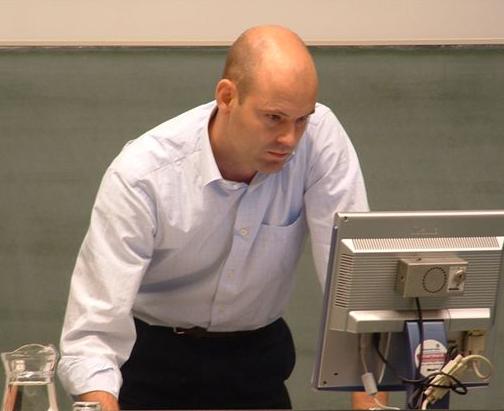}}]{Marcel Vos} is a researcher of the Spanish research council CSIC funded under the Ram\'on y Cajal program. He is based at the Instituto de F\'isica Corpuscular (IFIC, UVEG/CSIC) in Valencia, where he works on instrumentation for future collider experiments and searches for signatures of physics beyond the Standard Model in top quark production at the LHC. 
\end{IEEEbiography}


\vfill

\bibliographystyle{IEEEtran}
\bibliography{depfet}{}
\vfill
\end{document}